\pdfoutput=1

\documentclass[11pt]{article}

\usepackage[final]{acl}
\usepackage{booktabs}

\usepackage{times}
\usepackage{latexsym}
\usepackage{stfloats}
\usepackage{float}
\usepackage{microtype}
\usepackage{placeins}
\usepackage{amsfonts}

\usepackage[T1]{fontenc}

\usepackage[utf8]{inputenc}

\usepackage{microtype}
\usepackage{multirow}
\usepackage{natbib}

\usepackage{inconsolata}

\usepackage{graphicx}
 \usepackage{amsmath}

\def\code#1{\texttt{#1}}

\usepackage{newfloat}
\usepackage{booktabs}
\usepackage{float} 
\DeclareFloatingEnvironment[
    fileext=loa,
    listname=List of Prompts,
    name=Prompt,
    placement=tbhp
]{prompt}

\title{Crossing Domains without Labels: Distant Supervision for Term Extraction}

\author{Elena Senger$^{1,2}$ \quad Yuri Campbell$^{2}$ \quad Rob van der Goot$^{3}$ \quad Barbara Plank$^{1}$ \\[1.5ex]
$^1$MaiNLP, Center for Information and Language Processing, LMU Munich, Germany \\
$^2$Fraunhofer Center for International Management and Knowledge Economy IMW, Germany \\
$^3$Department of Computer Science, IT University of Copenhagen, Denmark \\
\texttt{elena.senger@cis.lmu.de, yuri.campbell@imw.fraunhofer.de} \\
\texttt{robv@itu.dk, b.plank@lmu.de}
}

\begin{document}
\maketitle
\begin{abstract}

Automatic Term Extraction (ATE) is a critical component in downstream NLP tasks such as document tagging, ontology construction and patent analysis. 
Current state-of-the-art methods require expensive human annotation and struggle with domain transfer, limiting their practical deployment. 
This highlights the need for more robust, scalable solutions and realistic evaluation settings. To address this, we introduce a comprehensive benchmark spanning seven diverse domains, enabling performance evaluation at both the document- and corpus-levels. 
Furthermore, we propose a robust LLM-based model that outperforms both supervised cross-domain encoder models and few-shot learning baselines and performs competitively with its GPT-4o teacher on this benchmark.
The first step of our approach is generating pseudo-labels with this black-box LLM on general and scientific domains to ensure generalizability. Building on this data, we fine-tune the first LLMs for ATE.  To further enhance document-level consistency, oftentimes needed for downstream tasks, we introduce lightweight post-hoc heuristics. Our approach exceeds previous approaches on 5/7 domains with an average improvement of 10 percentage points. We release our dataset and fine-tuned models to support future research in this area.\footnote{Dataset: \url{https://huggingface.co/datasets/ElenaSenger/SynTerm}; Model: \url{https://huggingface.co/ElenaSenger/DiSTER-Llama-3-8B-Instruct}}

\end{abstract}
\section{Introduction}

Automatic Term Extraction (ATE) is a crucial component of many NLP systems, with applications in information retrieval, machine translation, topic detection, and sentiment analysis \citep{tran2023recentadvancesautomaticterm, Xu.2025}. Traditional rule-based or frequency-based ATE systems, as well as state-of-the-art (SOTA) methods with pretrained models, rely heavily on fine-tuning with human-annotated datasets, which are typically available for only a handful of domains. Recent surveys explicitly highlight this dependence as a key limitation, noting that multi-domain ATE scenarios remain an unsolved challenge for current SOTA approaches \citep{tran2023recentadvancesautomaticterm, Xu.2025}. Large language models (LLMs), with their massive pretraining across diverse corpora, offer a promising path toward generalizable ATE. Yet, early applications of LLMs in term extraction remain limited and typically perform worse compared to supervised SOTA methods \citep{tran2023recentadvancesautomaticterm}. Moreover, proprietary black-box LLMs incur high API costs and pose privacy risks when handling sensitive or confidential data.

To address these limitations, we introduce a novel ATE framework: DiSTER (\textbf{Di}stant \textbf{S}upervision for \textbf{T}erm \textbf{E}xtraction with \textbf{R}obustness), that leverages LLMs with distant supervision. Our approach trains smaller, open models using synthetic data generated via pseudo-labels from a black-box LLM, thereby removing the need for human annotation and enabling cross-domain scalability. To enhance consistency within and across documents, we incorporate simple post-hoc consistency heuristics. These heuristics significantly improve F1 scores and oftentimes lead to more balanced precision and recall.

Moreover,  we perform a comprehensive empirical study, spanning the seven following domains: biomedicine, corruption, dressage, heart failure, coastal geography, computational linguistics and wind energy. Combining these established datasets makes this the largest and most diverse multi-domain ATE evaluation to date. We assess models under both corpus-level and document-level setups to better reflect real-world extraction needs. Our results demonstrate that training on distantly supervised data leads to notable improvements in cross-domain robustness and that post-hoc consistency enforcement yields further gains, boosting document-level F1 scores by up to 55 percentage points.
Our contributions are: 
\begin{itemize} 
\item We propose DiSTER, a novel distantly supervised ATE framework that combines synthetic data generation, LLM fine-tuning, and lightweight post-hoc consistency heuristics for robust and scalable term extraction without human annotation.
\item We demonstrate that a strategically constructed, domain-diverse synthetic training corpus significantly enhances cross-domain generalization.
\item We conduct the most comprehensive cross-domain ATE evaluation to date, spanning seven diverse domains and evaluating both corpus-level and document-level performance.
\end{itemize}

\section{Related work}

\subsection{Automatic Term Extraction}

Traditional ATE methods typically follow a two-step pipeline: (1) extracting candidate terms using linguistic and statistical features, and (2) ranking them based on termhood and unithood scores \citep{Xu.2025}.  Supervised machine learning approaches enhance this process using manually designed features and classifiers like SVMs, Random Forests, or CRFs \citep{tran2023recentadvancesautomaticterm}. With deep learning, models such as BiLSTMs, CNNs, and Transformers have been adopted for token classification and embedding-based approaches, showing SOTA results across languages and domains~\citep{tran2023recentadvancesautomaticterm, tran.2022, tran-2022-ensemble}. Recently, LLMs have entered the field. \citet{giguere-2023-leveraging} showed that GPT-4 \citep{openai2024gpt4technicalreport} performs well in zero-shot settings across legal, technical, and medical domains, outperforming statistical baselines on small test sets. Meanwhile, \citet{tran.2024} explored few-shot prompting with \code{LLaMA} and \code{GPT-3.5-Turbo} for the ACTER heartfailure dataset, though results still lag behind cross-domain sequence labeling with XLM-R \citep{conneau-etal-2020-unsupervised}.

\begin{figure*}
  \centering
  \includegraphics[width=1.0\linewidth]{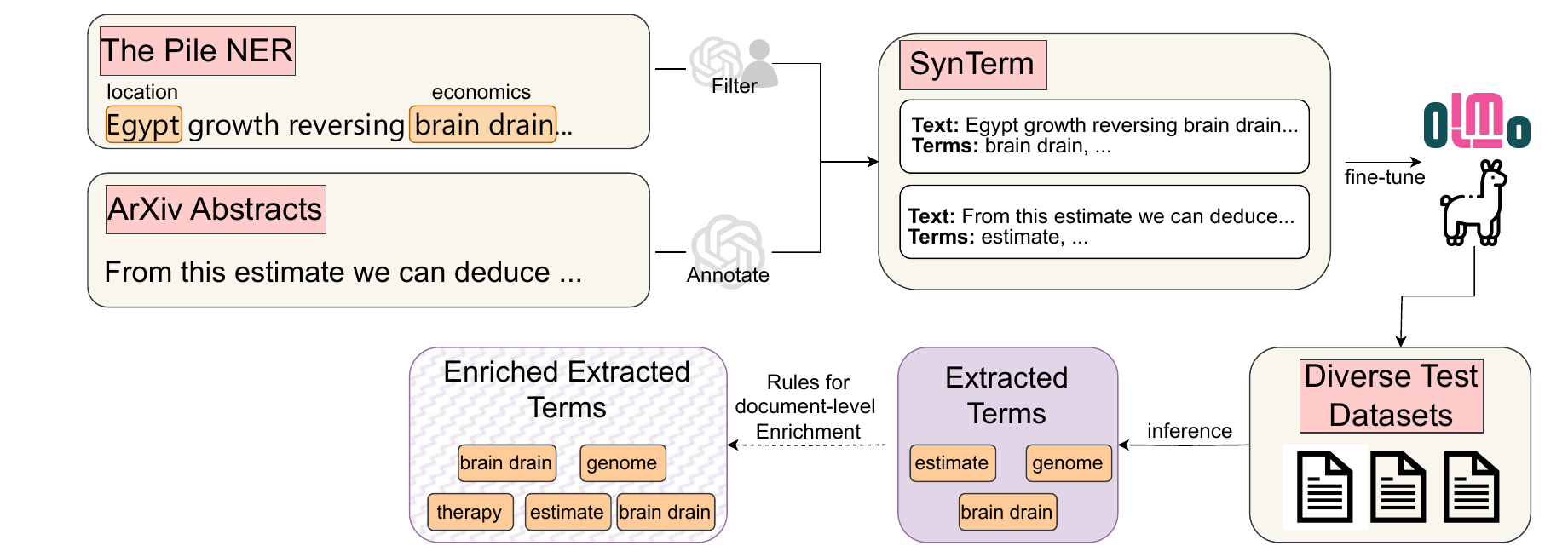}
  \caption{An overview of the key components of our DiSTER approach.}
  \label{fig:approach}
\end{figure*}
\subsection{Distillation and Pseudo-Labeling}

Knowledge Distillation (KD) transfers knowledge from larger teacher models to smaller student models.
In black-box KD, often used with proprietary LLMs, only the teacher’s outputs are available \citep{yang2024survey}, hence, the student model learns by mimicking the teacher’s generated sequences instead of its internal states.
Specifically, \textit{Labeling Knowledge}, when a LLM labels a set of examples based on an instruction (with or without demonstrations), is widely considered effective for transferring specific LLM skills \citep{xu2024survey}.
This approach has proven effective across diverse NLP tasks \citep{li2025learninglessknowledgedistillation}. 
In the information extraction domain, several studies have shown promising results: \citet{hsu2024leveraging} applied LLM weak supervision improving performance on medical entity extraction with minimal human annotation. Similarly, UniversalNER \citep{zhou2024universalnertargeteddistillationlarge} successfully distilled open-domain named-entity recognition (NER) capabilities from \code{GPT-3.5-Turbo-0301} into smaller models that ultimately outperformed their teacher. For more complex extraction tasks, MetaIE \citep{peng2024metaiedistillingmetamodel} employed distillation as a meta-learning framework to create task-flexible information extraction systems capable of adapting to various relation and entity types.

\section{DiSTER}

Our approach DiSTER first creates a distantly supervised dataset, then fine-tunes an LLM on that data to generate candidate terms, and lastly selectively applies post-hoc heuristics. This pipeline is illustrated in Figure \ref{fig:approach}.
We cover the details of each component in the following sections.

\subsection{Dataset Creation}

To create our synthetic dataset \textit{SynTerm} for model fine-tuning, we used the dataset of \citet{zhou2024universalnertargeteddistillationlarge} as a basis. 
Their synthetic NER dataset was generated by prompting \code{gpt-3.5-turbo-0301} to identify named entities within text snippets taken from \textit{The Pile} \citep{pile}.
 We utilize their annotations as a starting point, but apply a filtering step (as described next) to focus only on relevant entity types and added labeled data from \textit{arXiv} for broader domain coverage.

\paragraph{Entity Type Filtering}
In order to filter, two authors manually annotated the 130 most common entity types covering approximately 74\% of all extracted entities, labeling them as either \textit{terms} or \textit{non-terms} (e.g., person names, locations, etc.) following the ACTER annotation guidelines \citep{acter_annotation_guidelines}. This manual annotation process resulted in a Cohen’s kappa coefficient of 0.723, indicating substantial inter-annotator agreement. These human-provided labels were used for all entities within the 130 most frequent types. For the remaining 26\% of entities, which belong to less common types,  we relied on GPT-4o \citep{openai2024gpt4ocard} to classify whether phrases of a given entity type represent terms (see Appendix \ref{sec:type_prompt}). To assess consistency, we compared the model’s labels with the human annotations on the 130 most common types, which yielded a lower Cohen’s kappa of 0.45. Notably, the LLM tended to label fewer entity types as terms than the annotators (109 vs.\ 127 and 136, respectively). For instance, GPT-4o 
labeled entity types such as `event,' `nationality,' falsely as terms and instances of `medical conditions' as non-terms, whereas the annotators considered them valid terms. Overall, the potential noise introduced by the LLM is expected to be limited, as most extracted entities fall within the 130 most common types and were annotated manually.  Manual annotation of all entity types would have been prohibitively expensive, as the NER dataset contains a total of 13,020 entity types. By focusing on the most common types where noise reduction matters most, we ensured high-quality filtering for the majority of data while using GPT-4o only for the long-tail of rare entity types. 

\paragraph{Domain-Aware Data Augmentation}
To increase domain diversity, we synthesized labeled examples from two-sentence snippets from \textit{arXiv} abstracts using GPT-4o. 
The two-sentence snippets were chosen to closely match the length of data points from both \textit{The Pile} and the test datasets. As in the few-shot setup of \citet{tran.2024}, we included domain information in the extraction prompt (see Appendix \ref{sec:syn_data_prompt}). The domain information was derived from the \textit{arXiv} categories associated with each abstract. Given the domain-dependent nature of termhood \citep{Xu.2025}, we expect this to improve the relevance of extracted terms. We also incorporated domain labels into the final data points used for fine-tuning, an example is shown in Figure \ref{fig:conversation_example}. In contrast, for data points based on \textit{The Pile}, where no domain labels are available and thus not integrated in the data points, we prompted the model during training to extract terms without specifying a domain. 

In summary, we created two data subsets: \textit{TermPile} (45,432 instances), the filtered NER dataset based on \textit{The Pile} and, \textit{TermArXiv} (37,829 instances) the newly synthesized \textit{arXiv} data points. Combining the two yields our final \textit{SynTerm} dataset.

\begin{figure}
  \centering
  \colorlet{shadecolor}{blue!5}
  \fcolorbox{blue!40!black}{shadecolor}{
    \begin{minipage}{\columnwidth}
      \small
      \textbf{Human:} \\
      \textit{Text: From this estimate we can deduce that the spatially homogeneous Boltzmann equation is well posed in a class of measure-valued processes. We also prove in an appendix a basic lemma on the total variation of time-integrals of time-dependent signed measures.}
      
      \vspace{0.3em}
      
      \textbf{GPT:} \\
      I've read this text.
      
      \vspace{0.3em}
      
      \textbf{Human:} \\
      Please extract the terms from the text that are relevant to the domain: Probability, Analysis of PDEs.
      
      \vspace{0.3em}
      
      \textbf{GPT:} \\
      \textit{["spatially homogeneous Boltzmann equation", "measure-valued processes", "total variation", "time-integrals", "time-dependent signed measures"]}
    \end{minipage}
  }
  \caption{Conversation example showing extraction of domain-specific terms from an \textit{arXiv} text. For data points without specific domain, like the ones coming from the Pile, we substitute the domain by ``General''.}
  \label{fig:conversation_example}
\end{figure}

\begin{table*}
\centering
\scriptsize
\begin{tabular}{ll}
\toprule
\textbf{Domain} & \textbf{Example Sentence} \\
\midrule
\textit{corp} (Corruption) & The first criterion creates a link between the \textbf{offence} and the \textbf{legal person}. \\
\textit{equi} (Dressage / Equity) & They might go from a \textbf{lengthened stride} and \textbf{half halt} back to a \textbf{working trot}. \\
\textit{htfl} (Heart Failure) & \textbf{Heart failure} risk among \textbf{patients} with \textbf{rheumatoid arthritis} starting a \textbf{TNF antagonist}. \\
\textit{wind} (Wind Energy) & \textbf{Wind turbine technology} has developed rapidly in recent years and Europe is at the hub of this hightech \textbf{industry}. \\
\textit{coast} (Coastal Science) & \textbf{Coastal communities} are prone to a \textbf{natural disaster} such as \textbf{tsunami}. \\
\textit{genia} (Biomedical) & \textbf{HB24} is likely to have an important role in \textbf{lymphocytes} as well as in certain developing tissues . \\
\textit{acl} (Computational Linguistics) & \textbf{Word Identification} has been an important and active issue in \textbf{Chinese Natural Language Processing}. \\
\bottomrule
\end{tabular}
\caption{Example sentences from each of the seven domains used in our experiments. Terms are bold.}
\label{tab:domain_examples}
\label{tab:datasets}
\end{table*}

\subsection{LLM Fine-Tuning}
\label{sec:llm-fine-tuning}
After constructing the dataset for fine-tuning we perform standard instruction tuning in order to transfer the ATE skill to a smaller model.
Precisely, we fine-tune two smaller instruction tuned models, Llama-3-8B-Instruct (\code{LLaMA}) and Olmo-7B-Instruct (\code{Olmo}), with standard next-token prediction objective and conversation-style chat templates.\footnote{LLaMA model: \url{https://huggingface.co/meta-llama/Meta-Llama-3-8B-Instruct}, Olmo model: \url{https://huggingface.co/allenai/OLMo-7B-Instruct}}
In both cases, we use only completions in order to compute the loss, that is, only the tokens generated by the language model after the last ``Assistant''-marker.
The models were trained for $3$ epochs, with learning rate of $2e-4$ and batch size of $8$.
Only the final checkpoints were taken for further analysis.

\subsection{Post-hoc Consistency Enforcement}
 
To address the inconsistent extraction behavior of LLMs we introduce two lightweight post-hoc heuristics for enforcing consistency. The first, document-level consistency (DC) enforcement, aims to correct the LLM's tendency to return only one instance of each extracted term per document, even when multiple mentions occur. To remedy this, we identify all exact string matches of each LLM-extracted term within the document. This approach is conceptually aligned with prior work in NER that enforces intra-document label agreement for repeated spans \citep{ krishnan-manning-2006-effective,ijcai2020p550}. 

The second rule, corpus-level consistency (CC) enforcement, promotes any term extracted in at least 50\% of the documents it appears in to all such documents—addressing inconsistencies in LLM output. This simple heuristic mirrors the Term Re-extraction Model (TREM) by \citet{vu-etal-2008-term}, which reintroduced globally validated terms into individual documents, and aligns with frequency-based termhood estimation methods \citep{Kageura.1996}.  Such reinforcement mechanisms help align model outputs more closely with span-level gold annotations and improve both recall and consistency in document-level evaluations.

\section{SOTA Approaches}
To better evaluate DiSTER, we compare it against two strong baseline approaches that represent the current SOTA methodologies in ATE. 

\subsection{Sequence Labeling Approach}
We adopt the approach introduced by \citet{tran.2022}, which frames ATE as a sequence classification task using the IOB tagging scheme. This approach employs an \code{XLM-R-based} token classifier with standard hyper-parameters and has been shown to achieve SOTA in term extraction. It remains a strong baseline, as even recent few-shot LLM-based methods could not consistently outperform it \citep{tran.2024}.

\subsection{Few-Shot Approach}
We re-implement the few-shot in-context learning approach of~\citet{tran.2024}, but use cross-domain few-shot samples instead of samples from the target test data.
Following \citet{tran.2024}, each prompt contains three examples. For every target test domain, we select examples showing the highest semantic similarity to the domain name, as measured by the \code{paraphrase-multilingual-mpnet-base-v2} embedding model.\footnote{https://huggingface.co/sentence-transformers/paraphrase-multilingual-mpnet-base-v2} Each example is structured as a demonstration pairing a source sentence with its corresponding extracted terms. In alignment with \citet{tran.2024}, we employ Direct Term Extraction rather than IOB tagging, the LLM  explicitly outputs the identified terms.
We enriched the prompt templates for extraction as introduced in \citet{tran.2024}, with complete specifications of these templates provided in Appendix \ref{app:few-shot-template}.

\section{Benchmark Datasets}

We make use of several established benchmarks, resulting in seven datasets. They are topically quite distant, which leads to substantially different types of terms, often including domain-specific jargon (see Table \ref{tab:datasets}). All datasets have predefined splits, except for the ACTER datasets, which were introduced in a cross-domain setting. While most original studies report in-domain performance, we only use cross-domain test splits in this work.

We employ the ACTER dataset introduced by the TermEval 2020 Shared Task~\citep{rigouts-terryn-etal-2020-termeval}, which includes four subsets with each a different domain: heart failure (\textit{htfl}), wind energy (\textit{wind}), dressage (equity) (\textit{equi}), and corruption (\textit{corp}). Secondly, we use the CoastTerm (\textit{coast}) dataset~\citep{delaunay2024coasttermcorpusmultidisciplinaryterm} consisting of scientific abstracts focused on coastal regions. Due to the inherently interdisciplinary nature of coastal studies, the texts include a wide range of specialized terms spanning domains such as environmental science, geography, ecology, and sociology. We also incorporate the \textit{genia} dataset, a standard benchmark for biomedical term extraction \citep{kim-etal-2011-overview-genia}. Finally, we use the ACL-RD-TEC 2.0 \textit{acl} dataset, which contains abstracts from the ACL Anthology from the domain of \textit{computational linguistics} \citep{qasemizadeh-schumann-2016-acl}.

\begin{table*}
\centering
\scriptsize
\begin{tabular}{l|c|c|c|c|c|c|c|c}
\toprule
\textbf{Model} & \textbf{\textit{corp} F1} & \textbf{\textit{equi} F1} & \textbf{\textit{htfl} F1} & \textbf{\textit{wind} F1} & \textbf{\textit{coast} F1} & \textbf{\textit{genia} F1} & \textbf{\textit{acl} F1} & \textbf{Avg F1} \\
\midrule
IOB sequence labeling (cross-domain) & 31.35 & 41.64 & 14.60 & \textbf{44.26} & 14.14 & 18.20 & 10.55 & 24.96 \\
IOB sequence labeling \textit{SynTerm} & 29.89 & 32.52 & 42.74 & 31.17 & 56.49 & 42.33 & 56.78 & 41.70 \\
\midrule
\code{LLaMA} few-shot (cross domain) & 10.47 & 41.74 & \textbf{51.29} & 33.39 & 42.71 & 45.33 & 40.93 & 37.98 \\
\code{LLaMA} few-shot \textit{SynTerm} & \underline{32.64} & \underline{43.53} & 41.71 & 38.47 & 36.67 & \underline{46.99} & 51.62 & 41.94 \\
\code{Olmo} few-shot (cross domain) & 27.09 & 37.13 & 49.87 & 34.54 & 49.00 & 46.40 & 31.23 & 39.32 \\
\midrule
\code{LLaMA} fine-tuned \textit{SynTerm} (DiSTER) & \textbf{37.93} & \textbf{45.54} & \underline{50.10} & \underline{42.93} & \textbf{66.96} & \textbf{51.80} & \textbf{65.37} & \textbf{51.52} \\

\code{Olmo} fine-tuned \textit{SynTerm} (DiSTER) & 32.03 & 34.37 & 45.10 & 35.28 & \underline{57.48} & 44.29 & \underline{57.73} & \underline{43.75} \\
\midrule
\midrule
Teacher Model zero-shot & 38.18 & 50.96 & 51.54 & 45.83 & 62.58 & 50.19 & 62.66 & 51.70 \\
Teacher Model few-shot (cross domain) & 41.19 & 40.66 & 53.57 & 36.40 & 62.89 & 48.21 & 60.82 & 49.10 \\
\bottomrule
\end{tabular}
\caption{Corpus-level F1 scores across seven datasets. Best result per dataset is marked in bold, second best results are underlined. Teacher model performance is included for comparison.}
\label{tab:f1_corpus}
\end{table*}

\section{Experimental Setup}

\subsection{Evaluation Strategies}
We employ both \textit{corpus-level} and \textit{document-level} evaluation. The \textit{corpus-level} approach aggregates predictions and gold annotations across the entire dataset before computing metrics, while the \textit{document-level} strategy calculates metrics per document and then averages results. 

\subsection{Model Configurations}
We evaluate three distinct model categories. First, for our method, DiSTER, which relies on \textbf{fine-tuned models}, we use
 two instruction tuned LLMs on our synthetic data: Llama-3-8B-Instruct (\code{LLaMA}) and Olmo-7B-Instruct (\code{Olmo}). 
Regarding the \textbf{few-shot prompted models}, 
we evaluate the same LLM architectures (\code{LLaMA} and \code{Olmo}) in a few-shot setting, using cross-domain demonstrations from the remaining datasets. For each target domain, we construct prompts with semantically similar examples from other domains. Additionally, we evaluate the better-performing model \code{LLaMA} using demonstrations from our dataset \textit{SynTerm}.
For the \textbf{IOB sequence labeling approaches}, we implement two training configurations. Firstly, we implement a leave-one-out approach where for each test domain, we train on five domains and validate on the sixth. We consistently use \textit{wind} for validation (given its STEM domain alignment) except when testing on \textit{wind} itself, where we validate on \textit{htfl}. In our second configuration, we train on our \textit{SynTerm} dataset to enable direct comparison with the fine-tuned LLMs (the full DiSTER methodology).

This experimental design allows us to systematically evaluate the impact of model architecture, training methodology, and data composition on cross-domain generalization in ATE.

\section{Results}
\label{sec:results}
\begin{table}
    \centering
    \small
    \setlength{\tabcolsep}{4pt} 
    \begin{tabular}{lcccc}
        \toprule
        \textbf{Val/Test} & \textbf{\textit{corp} F1} & \textbf{\textit{equi} F1} & \textbf{\textit{htfl} F1} & \textbf{\textit{wind} F1} \\
        \midrule
        \textit{corp}  & --     & 45.32 & 40.98 & 33.74 \\
        \textit{equi}  & 6.10   & --    & 46.68 & 30.78 \\
        \textit{htfl}  & 6.40   & 51.80* & --    & 42.57 \\
        \textit{wind}  & 7.89  & 31.57 & 31.68 & --    \\
        \bottomrule
    \end{tabular}
    \caption{\small  F1 scores across ACTER domains using a leave-one-out setup: one domain for testing, one for validation, and the remaining two for training. \*The \textit{htfl} score is reproduced from \citet{tran.2022} under the original Shared Task setting. }
    \label{tab:IOB}
\end{table}

\begin{table*}
\centering
\scriptsize
\begin{tabular}{l|c|c|c|c|c|c|c|c}
\toprule
\textbf{Model} 
  & \textbf{\textit{corp} F1} 
  & \textbf{\textit{equi} F1} 
  & \textbf{\textit{htfl} F1} 
  & \textbf{\textit{wind} F1} 
  & \textbf{\textit{coast} F1} 
  & \textbf{\textit{genia} F1} 
  & \textbf{\textit{acl} F1} 
  & \textbf{Avg F1} \\
\midrule
IOB sequence labeling (cross-domain) 
  & 25.66 & 46.91 & 8.01  & \textbf{49.90} & 11.41 & 13.49 & 8.72  & 23.44 \\
IOB sequence labeling \textit{SynTerm} 
  & 21.65 & 33.81 & 39.63 & 26.52 & 42.11 & 28.27 & 39.37 & 33.05 \\
\midrule
\code{LLaMA} few-shot (cross domain)
  & 5.82  & \textbf{58.50} & 43.65 & 37.49 & 35.41 & 42.88 & 32.05 & 36.54 \\
  \code{LLaMA} few-shot \textit{SynTerm} & 29.22 & 38.57 &  25.23 & 41.25 & 30.23 & 45.77 & 42.20 & 36.07 \\
\code{Olmo} few-shot (cross domain)
  & 14.52 & 48.41 & 47.28 & 37.94 & 44.14 & 44.44 & 23.28 & 37.14 \\
\midrule
\code{LLaMA} fine-tuned \textit{SynTerm}
  & \textbf{38.25} & \underline{51.60} & \textbf{49.03} & \underline{49.87} & \textbf{63.19} & \textbf{51.10} & \textbf{56.31} & \textbf{51.34} \\
\code{Olmo} fine-tuned \textit{SynTerm}
  & 35.63 & 43.77 & \underline{47.51} & 39.12 & 55.60 & 38.55 & 50.07 & 44.32 \\
\bottomrule
\end{tabular}
\caption{Document-level F1 scores across seven datasets. Best result per dataset is marked in bold, second best results are underlined.}
\label{tab:f1_document}
\end{table*}

\begin{table*}
\centering
\resizebox{2\columnwidth}{!}{
\begin{tabular}{l|cc|cc|cc|cc|cc|cc|cc|c|c}
\toprule
\textbf{Model} & 
\multicolumn{2}{c|}{\textit{corp}} & 
\multicolumn{2}{c|}{\textit{equi}} & 
\multicolumn{2}{c|}{\textit{htfl}} & 
\multicolumn{2}{c|}{\textit{wind}} & 
\multicolumn{2}{c|}{\textit{coast}} & 
\multicolumn{2}{c|}{\textit{genia}} & 
\multicolumn{2}{c|}{\textit{acl}} & 
\textbf{Mean F1} & \textbf{Mean $|\text{R-P}|$} \\
& F1 & $|\text{R-P}|$ & F1 & $|\text{R-P}|$ & F1 & $|\text{R-P}|$ & F1 & $|\text{R-P}|$ & F1 & $|\text{R-P}|$ & F1 & $|\text{R-P}|$ & F1 & $|\text{R-P}|$  & (\%) & (\%) \\
\midrule
\code{LLaMA} fine-tuned & 38.25 & 7.74 & 51.60 & 15.33 & 49.03 & 10.27 & 49.87 & 4.11 & 63.19 & 19.92 & 51.10 & 18.55 & 56.31 & 22.53 & 51.91 & 13.63 \\
\code{LLaMA} fine-tuned DC & 39.87 & 10.78 & 54.27 & 12.77 & 51.36 & 8.21 & 53.06 & 9.01 & 64.25 & 18.73 & 52.87 & 21.88 & \textbf{65.14} & 12.72 & 54.40 & 13.73 \\
\code{LLaMA} fine-tuned CC & 40.05 & 14.89 & 54.79 & 5.68 & 52.01 & 1.92 & 51.30 & 11.89 & 63.74 & 12.47 & 51.76 & 27.51 & 55.98 & 17.84 & 52.80 & 13.74 \\
\code{LLaMA} fine-tuned DC + CC & \textbf{41.25} & 19.32 & \textbf{57.14} & 1.82 & \textbf{54.22} & 0.80 & \textbf{54.55} & 18.68 & \textbf{64.53} & 10.72 & \textbf{53.74} & 32.00 & 63.98 & 5.40 & \textbf{55.92} & 12.68 \\
\bottomrule
\end{tabular}
}
\caption{F1 scores (bolded for highest score per dataset), absolute precision-recall gaps, and mean F1 score across datasets. Showing the influence of the document consistency (DC) and corpus consistency (CC) heuristics. }
\label{tab:doc_level_rules}
\end{table*}
Table~\ref{tab:f1_corpus} presents the corpus-level F1 scores.
 Notably, our models achieve the highest F1 scores in most domains, surpassing both sequence-labeling and few-shot prompting methods. The fine-tuned \code{LLaMA} model reaches the best overall performance.
\code{Olmo} also performs consistently well across domains, demonstrating the effectiveness of our approach even for open-data models. While few-shot prompted models show competitive performance in specific domains (e.g., \code{LLaMA} on \textit{htfl}), their performance remains fundamentally inconsistent, with significant variability across different domains such as \textit{corp}. Using our \textit{SynTerm} dataset for demonstrations yields a four percentage point average improvement compared to cross-domain few-shot with LLaMA. Notably, the performance gains for \textit{corp} and \textit{wind}, the two domains with the lowest cross-domain few-shot scores, suggest that \textit{SynTerm} helps achieve more robust overall few-shot performance.
The supervised cross-domain IOB models reveal inherent limitations, particularly struggling with recall under domain shift. Even when trained on our \textit{SynTerm} dataset, these models consistently underperform compared to our fine-tuned LLM-based approaches.
Remarkably, Table~\ref{tab:f1_corpus} also shows that the fine‑tuned \code{LLaMA} student surpasses the teacher in corpus‑level F1 on three of seven test corpora and stays on-par on the macro average.
These findings provide strong support for the generalizability and use of DiSTER as a pivotal strategy for the development of more domain-robust and light-weight ATE systems.

\section{Analysis}



In this section, we begin by comparing the student to its black‑box teacher. Then we inspect the error sources causing the few-shot and IOB model failure. Finally, we present a discussion about the distinctions and use cases of document-level evaluation and the impact of our post‑hoc heuristics.

\subsection{The Student Rivals Its Teacher}\label{subsec:student-vs-teacher}


Table~\ref{tab:f1_corpus} shows that DiSTER effectively distills ATE capabilities from the teacher into the much lighter \code{LLaMA} student. The student outperforms the teacher on three of seven domains and comes within 1.5~F1 points on two more, demonstrating strong competitive performance despite its smaller size. The largest gains appear in domains semantically aligned with the synthetic corpus (\textit{acl}, \textit{coast}, \textit{genia}). Where overlap is weaker, the student tends to lag further behind, suggesting that broader domain coverage in the pseudo-labeled training data could close the remaining gaps. See Appendix \S\ref{appendix:overlap-analysis} for an analysis of training data overlap.

We hypothesize that distillation works due to two complementary factors: (i) sequence-level supervision encourages the student to mimic the teacher’s span predictions exactly, reducing prompt sensitivity and reinforcing task-specific patterns; and (ii) domain cues in the synthetic data act as scaffolding: smaller models benefit from domain-specific regularities, aiding generalization in overlapping domains. This suggests that strong cross-domain performance still depends on diverse fine-tuning data, making automated labeling approaches like DiSTER a cost-effective path to improving ATE generalizability in smaller, deployable models.



\subsection{Where IOB and Few-Shot Fail}
\label{sec:length-counts}

Two systematic error sources (\emph{extraction count} and \emph{term-span length}) explain much of the under-performance observed in the baseline systems, as thoroughly discussed in Appendix \ref{appendix:term-length-extraction-counts}. In brief, in the few-shot setting, models often return few or no candidates. On the \textit{corp} subset, the median number of predicted terms is zero. Even when terms are extracted, their median length far exceeds the gold standard, which severely depresses recall (Table~\ref{tab:combined-median-term-stats}).

The supervised IOB model exhibits the converse pathology. When trained on \textit{SynTerm}, the model assigns the term label to overly long spans that often include stop words, thereby inflating recall while harming precision.
DiSTER’s LLM fine-tuning better addresses this challenge by producing extraction counts and term spans that are closer to the gold standard.
Therefore, we posit that the underlying limitation is architectural. These findings are consistent with the instability patterns observed under cross-domain training (Table~\ref{tab:IOB}; see Appendix~\ref{appenidx:instability} for a detailed analysis). Even within the original four ACTER domains, F1 scores can drop by over 30 points depending on the validation split. Together, these analyses highlight the need for a more flexible architecture and targeted fine-tuning to best leverage the distillation data \textit{SynTerm} provides.


\subsection{Document-Level Evaluation}

While corpus-level ATE has been the primary focus in prior evaluations, document-level evaluation better reflects downstream tasks like computer-assisted translation and information retrieval \citep{sajatovic-etal-2019-evaluating}. As shown in Table~\ref{tab:f1_document}, document-level F1 scores are generally lower, reflecting the greater challenge of consistently extracting terms within individual documents. In particular, recall is significantly lower and, in most cases, falls below precision, reversing the trend observed at the corpus-level (see Appendix~\ref{appendix:precision_recall}). At the corpus level, terms from all documents are pooled, so consistency within individual documents or term repetition matters less.

As shown in Table~\ref{tab:doc_level_rules}, applying consistency enforcement heuristics improves document-level F1 scores. Especially, the \textit{acl} dataset, with many term repetitions per document, sees a nine-point F1 gain. Since the heuristics target different patterns, within-document (DC) and cross-corpus (CC) term repetition, their effects are complementary.  When combined, they yield the highest F1 scores in six of seven evaluated datasets. Additionally, when precision exceeds recall, these heuristics narrow the precision-recall gap, resulting in a more balanced performance.

\section{Conclusion}
We introduced DiSTER, a scalable and robust ATE framework combining synthetic data generation, LLM fine-tuning, and post-hoc consistency heuristics. By using pseudo-labels from a black-box LLM, we built the diverse \textit{SynTerm} corpus to support cross-domain generalization. The fine-tuned LLMs, especially \code{LLaMA}, outperform both supervised sequence labeling and few-shot prompting and perform competitively with the GPT-4o teacher model, despite the size gap. Our results highlight the importance of data composition in cross-domain ATE and show that our approach generalizes well even to less related domains.  We also show that document-level evaluation reveals important limitations in consistency, which can be effectively addressed using simple heuristics.  Crucially, DiSTER eliminates the need for domain-specific training, making ATE more scalable and practical. 
%
%


\section*{Limitations}

Our approach, while effective, has several limitations.
First, the pseudo-labels used for training are derived from a black-box LLM and may contain noise, especially for rare or ambiguous terms. Second, while our models perform well even in domains with limited overlap in the training data, generalization to entirely unseen or underrepresented domains cannot be guaranteed. Expanding the diversity of synthetic data could further strengthen cross-domain robustness.  Third, the post-hoc consistency heuristics are simple heuristics and do not handle paraphrases or semantic variants, which could limit precision. Fourth, while we use relatively lightweight LLMs, fine-tuning still demands substantial computational resources, potentially limiting accessibility for low-resource settings.
Lastly, our experiments are conducted only in English, and computational cost of fine-tuning open-weight models may hinder adoption in low-resource settings.

\newpage
\bibliography{custom}

\appendix

\section{Use Of AI Assistants}
The authors acknowledge the use of ChatGPT and Claude solely for correcting grammatical and spelling errors, and providing assistance with coding.

\section{Few-shot prompt template}
\label{app:few-shot-template}
Extending the few-shot prompt templates introduced by \citep{tran.2024},
we design two modifications: 
\begin{itemize}
    \item \textbf{Context Enrichment (CE)}: At the beginning of the user prompt, we add the ISO definition of a \textit{term} --- \textit{Terms are “the designation of a defined concept in a special language by a linguistic expression.” (ISO 1087). A term is a word or a phrase that has a specific meaning in a particular context/domain, such as a scientific term or a technical concept.}
    \item \textbf{Assistance Response Guidance (ARG)}: Instead of freely letting the language model begin the assistant's response, we prepend to the to the response the sentence: \textit{I have extracted the terms from the text. Here is the list of terms:}, and let the LLM complete it.
\end{itemize}
While CE aims to provide the LLM with more knowledge about the task at hands, ARG deals with the inherent structure-free and stochastic nature of LLMs' generations, in order to facilitate parsing.

\section{Prompt for Entity Type Labeling}
\label{sec:type_prompt}

Prompts \ref{prompt:type-system-prompt} and \ref{prompt:type-user-prompt} show the system and user prompts used for entity-type-label generation, respectively.

\begin{prompt}
\centering
\begin{tabular}{|p{0.95\linewidth}|}
\hline
\textbf{System Prompt} \\
\hline
You are a terminology research expert. Your task is to help the user by answering the following question.\\
\hline
\end{tabular}
\caption{System prompt used for entity-type-label generation. \label{prompt:type-system-prompt}}
\end{prompt}

\begin{prompt}
\centering
\begin{tabular}{|p{0.95\linewidth}|}
\hline
\textbf{User Prompt} \\
\hline
\#\#\# Context \\
    - Terms are “the designation of a defined concept in a special language by a linguistic expression.” (ISO 1087).\\
    - A term is a word or a phrase that has a specific meaning in a particular context, such as a scientific term or a technical concept.\\
    - The goal is the identify domain-specific concepts not named entities.\\
    - A numerical value or date is not considered a term.\\
    - A organization, group or person is not considered a term.\\
    - Any non-scientific content such as websites, URLs, email addresses, HTML tags, code snippets, etc, are not considered terms.\\
    - A location (country, state, place, ...) is not considered a term.\\

    \#\#\# Instructions \\
    - You will be given a question about a type of expressions in natural language. \\
    - Your task is to determine whether the given expression is either:\\
        1) a term type, or\\
        0) no term in the sense of the definition above.\\
    - Do not overthink the question. Answer based on your intuition and knowledge.\\
    - If the context is not clear, use the most common interpretation of the expression type.\\
    - After you explain your answer, produce a json object with the following format: \texttt{\{ "answer": [Option Number] \}}, where [Option Number] is 0, 1, or 2.\\

    \#\#\# Question\\
    Given expressions of the type \{expression\_type\}, would you consider it\\
    - a term (1), or\\
    - no term (0)?\\
    
    """ \\
\hline
\end{tabular}
\caption{User prompt used for entity-type-label generation. \label{prompt:type-user-prompt}}
\end{prompt}

\section{Prompts for Data Synthetization}
\label{sec:syn_data_prompt}
Prompts \ref{prompt:synth-system-prompt} and \ref{prompt:synth-user-prompt} show the system and user prompts used for pseudo-label generation, respectively.

\begin{prompt}[h!]
\centering
\begin{tabular}{|p{0.95\linewidth}|}
\hline
\textbf{System Prompt} \\
\hline
You are a terminology research expert. Your task is to help the user extracting terms from scientific abstracts. \\
\hline
\end{tabular}
\caption{System prompt used for pseudo-label generation. \label{prompt:synth-system-prompt}}
\end{prompt}

\begin{prompt}[h!]
\centering
\begin{tabular}{|p{0.95\linewidth}|}
\hline
\textbf{User Prompt} \\
\hline
\#\#\# Context \\
- Terms are “the designation of a defined concept in a special language by a linguistic expression.” (ISO 1087).\\
- A term is a word or a phrase that has a specific meaning in a particular context/domain, such as a scientific term or a technical concept.\\
- Only extract terms that are relevant to the domain.\\
- The goal is to identify domain-specific concepts not named entities.\\
- A numerical value or date is not considered a term.\\
- An organization, group or person is not considered a term.\\
- A location (country, state, place, ...) is not considered a term.\\
\\
\#\#\# Instructions \\
- You will be given an excerpt from a scientific abstract.\\
- Your task is to extract all terms from the text.\\
- Please return only a comma separated list of correct extractions without any additional information.\\
- If there are no terms in the text, return an empty string.\\
\\
\#\#\# Question\\
The domain: \{domain\}\\
The abstract: \{abstract\}\\
\\
\#\# Return list of extracted terms:\\
\hline
\end{tabular}
\caption{User prompt used for pseudo-label generation. \label{prompt:synth-user-prompt}}
\end{prompt}


\section{Precision and Recall per Domain and Model}
\label{appendix:precision_recall}
The Table~\ref{tab:precision_recall_corpus} shows the corpus-level precision and recall per dataset and model, while Table ~\ref{tab:precision_recall_document} shows the document-level scores.
\begin{table*}
\resizebox{2\columnwidth}{!}{
\begin{tabular}{l|cc|cc|cc|cc|cc|cc|cc}
\toprule
\textbf{Model}
  & \multicolumn{2}{c|}{\textbf{acter\_corp}}
  & \multicolumn{2}{c|}{\textbf{acter\_equi}}
  & \multicolumn{2}{c|}{\textbf{acter\_htfl}}
  & \multicolumn{2}{c|}{\textbf{acter\_wind}}
  & \multicolumn{2}{c|}{\textbf{coast}}
  & \multicolumn{2}{c|}{\textbf{genia}}
  & \multicolumn{2}{c}{\textbf{acl}} \\
  & \textbf{P} & \textbf{R}
  & \textbf{P} & \textbf{R}
  & \textbf{P} & \textbf{R}
  & \textbf{P} & \textbf{R}
  & \textbf{P} & \textbf{R}
  & \textbf{P} & \textbf{R}
  & \textbf{P} & \textbf{R} \\
\midrule
IOB sequence labeling \textit{SynTerm} 
  & 19.2  & 67.52
  & 21.83 & 63.71
  & 30.4  & 69.86
  & 19.32 & 80.69
  & 49.04 & 66.5
  & 30.49 & 69.17
  & 47.75 & 70.02 \\
IOB sequence labeling (cross-domain) 
  & 43.5  & 24.5
  & 39.33 & 44.23
  & 36.27 & 9.14
  & 33.08 & 66.85
  & 56.02 & 8.09
  & 20.12 & 16.61
  & 68.42 & 5.72 \\

  \midrule
  \code{LLaMA} few-shot (cross domain)
  & 10.39 & 10.53
  & 28.85 & 74.83
  & 47.52 & 55.68
  & 21.85 & 70.62
  & 68.17 & 31.16
  & 36.49 & 59.81
  & 68.5  & 29.24 \\
  \code{LLaMA} few-shot \textit{SynTerm} 
 & 23.09 &  55.65   
 &  35.05 &  57.44 
   &  47.48  & 37.19 
   &  26.64 &  69.20
   &  72.55 &  24.54
   &  37.13 &  63.98
  & 71.20 & 40.48  \\
\code{Olmo} few-shot (cross domain)
  & 28.89 & 25.5
  & 24.74 & 74.41
  & 43.44 & 17.17
  & 23.26 & 67.14
  & 65.05 & 39.31
  & 41.51 & 1.7
  & 67.7  & 20.29 \\

  \midrule
\code{LLaMA} fine-tuned \textit{SynTerm}
  & 26.34 & 67.74
  & 33.81 & 69.73
  & 39.89 & 67.34
  & 29.82 & 76.55
  & 73.87 & 61.24
  & 40.86 & 70.75
  & 71.66 & 60.1 \\

\code{Olmo} fine-tuned \textit{SynTerm}
  & 22.08 & 58.31
  & 25.81 & 51.42
  & 35.34 & 62.31
  & 23.44 & 71.28
  & 61.71 & 53.8
  & 34.63 & 61.4
  & 65.21 & 51.78 \\
\bottomrule
\end{tabular}
}
\caption{Corpus-level evaluation: Precision (P) and Recall (R) scores for each dataset across models.}
\label{tab:precision_recall_corpus}
\end{table*}

\begin{table*}
\resizebox{2\columnwidth}{!}{
\begin{tabular}{l|cc|cc|cc|cc|cc|cc|cc}
\toprule
\textbf{Model}
  & \multicolumn{2}{c|}{\textbf{acter\_corp}}
  & \multicolumn{2}{c|}{\textbf{acter\_equi}}
  & \multicolumn{2}{c|}{\textbf{acter\_htfl}}
  & \multicolumn{2}{c|}{\textbf{acter\_wind}}
  & \multicolumn{2}{c|}{\textbf{coast}}
  & \multicolumn{2}{c|}{\textbf{genia}}
  & \multicolumn{2}{c}{\textbf{acl}} \\
  & \textbf{P} & \textbf{R}
  & \textbf{P} & \textbf{R}
  & \textbf{P} & \textbf{R}
  & \textbf{P} & \textbf{R}
  & \textbf{P} & \textbf{R}
  & \textbf{P} & \textbf{R}
  & \textbf{P} & \textbf{R} \\
\midrule
IOB sequence labeling \textit{SynTerm} 
  & 13.08 & 62.73
  & 23.15 & 62.65
  & 28.83 & 63.38
  & 16.42 & 68.77
  & 30.81 & 66.5
  & 18.23 & 62.88
  & 27.05 & 72.28 \\
IOB sequence labeling (cross-domain) 
  & 45.92 & 17.8
  & 67.51 & 35.95
  & 32.67 & 4.56
  & 46.83 & 53.98
  & 45.38 & 6.53
  & 12.37 & 14.84
  & 57    & 0.05 \\
  \midrule
\code{LLaMA} few-shot (cross domain)
  & 14.62 & 3.64
  & 57.58 & 59.44
  & 62.27 & 33.6
  & 34.15 & 41.54
  & 61.66 & 24.84
  & 39.45 & 46.95
  & 63.62 & 21.42 \\
  \code{LLaMA} few-shot \textit{SynTerm} 
  & 29.20  &  29.23 &
  57.91 &  28.92 &  
  56.08 & 16.27  &  
  40.92 &  41.58 &  
  69.78 & 19.29  &  
  39.27 &  54.85 &  
   66.84 & 30.84 \\
  \code{Olmo} few-shot (cross domain)
  & 28.93 & 9.69
  & 49.49 & 47.38
  & 56.42 & 40.69
  & 36.88 & 39.07
  & 61.01 & 34.58
  & 38.36 & 52.82
  & 62.35 & 14.32 \\
  \midrule
\code{LLaMA} fine-tuned \textit{SynTerm}
  & 34.77 & 42.51
  & 60.38 & 45.05
  & 54.7  & 44.43
  & 47.9  & 52.01
  & 74.68 & 54.76
  & 43.46 & 62.01
  & 69.75 & 47.22 \\
\code{Olmo} fine-tuned \textit{SynTerm}
  & 28.15 & 48.54
  & 49.13 & 39.46
  & 47.4  & 47.62
  & 32.03 & 50.25
  & 60.44 & 51.48
  & 28.54 & 59.37
  & 60.66 & 42.63 \\
\bottomrule
\end{tabular}
}
\caption{Document-level evaluation: Precision (P) and Recall (R) scores for each dataset across models.}
\label{tab:precision_recall_document}
\end{table*}

\newpage

\section{Qualitative Examples}
\label{appendix:qualitative_example}
Table~\ref{tab:predictions} presents a qualitative comparison of different models on the same input sentence. The extracted terms are highlighted in bold. This example underscores the contrast in recall capabilities between various approaches.

In particular, the IOB sequence labeling model trained on \textit{SynTerm} demonstrates high recall, but low precision. In contrast, the few-shot approaches  tend to miss key terms, reflecting their comparatively lower recall. Fine-tuned models  perform significantly better than their few-shot counterparts, aligning more closely with the gold terms.
\begin{table*}
\centering
\scriptsize
\begin{tabular}{ll}
    \toprule
    \textbf{Model} & \textbf{Text} \\
    \midrule
    IOB sequence labeling \textit{SynTerm} &
      This \textbf{is} due \textbf{to} the \textbf{fact} that \textbf{corruption}
      \textbf{is} often referred \textbf{to} \textbf{as} the \textbf{crime}
      without ( direct ) victim. \\
    IOB sequence labeling (cross-domain) &
      This is due to the fact that \textbf{corruption} is often referred to as the crime without ( direct ) victim. \\
      \midrule
    \code{Olmo} few-shot (cross domain) &
      This is due to the fact that corruption is often referred to as the \textbf{crime} without ( direct ) victim. \\
    \code{LLaMA} few-shot (cross domain) &
      This is due to the fact that corruption is often referred to as the crime without ( direct ) victim. \\
      \midrule
    \code{LLaMA} fine-tuned \textit{SynTerm} &
      This is due to the fact that \textbf{corruption} is often referred to as the \textbf{crime} without ( direct ) \textbf{victim}. \\
    \code{Olmo} fine-tuned \textit{SynTerm} &
      This is due to the fact that \textbf{corruption} is often referred to as the \textbf{crime} without ( direct ) victim. \\
    \bottomrule
  \end{tabular}
  \caption{Models and their extracted terms highlighted in the sentence. The gold-terms are : ['corruption', 'crime', 'victim'].}
  \label{tab:predictions}
\end{table*}


\section{Unique terms among datasets}
\label{appendix:term-set-overlap}

\begin{figure}
  \centering
\includegraphics[width=0.50\textwidth]{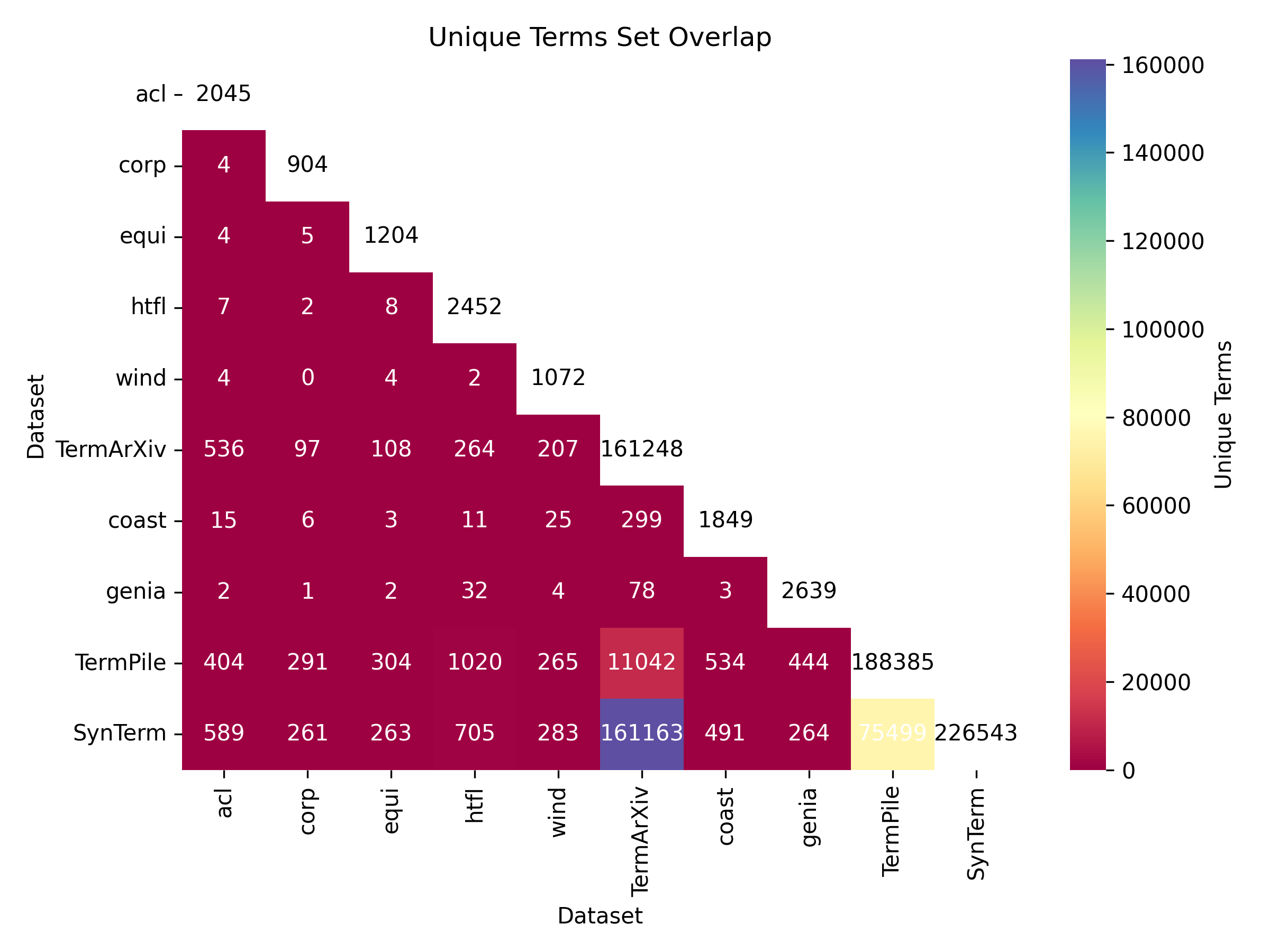}
  \caption{Counts of unique terms among the considered datasets. Off-diagonal counts represent common terms.}
  \label{fig:term-set-overlap-counts}
\end{figure}

Figure \ref{fig:term-set-overlap-counts} shows the amount of unique terms per dataset on the diagonal and the amount of unique common terms among the datasets on the lower triangular part.
We observe that, as expected, \textit{SynTerm} has the highest overlap with its source datasets, \textit{TermPile} and \textit{TermArXiv}.
Furthermore, \textit{SynTerm} shows considerable overlap with all other datasets considered.
Contrary to the expected, the overlap comes not only from \textit{TermPile}, but also from the \textit{TerArXiv} dataset, as it shows also common unique term counts in the same order of maginitude as \textit{TermPile}.
Interestingly, \textit{SynTerm} has almost all unique terms present in \textit{TermArXiv}, which speaks to the relatively restricted domain and language used in the original dataset.

\section{Term Length and Extraction Counts}
\label{appendix:term-length-extraction-counts}

\begin{table*}[h!]
\centering
\scriptsize
\setlength{\tabcolsep}{3.5pt}
  \begin{tabular}{lccccccccccccccc}
    \toprule
    \textbf{Model} & \multicolumn{2}{c}{\textit{corp}} & \multicolumn{2}{c}{\textit{equi}} & \multicolumn{2}{c}{\textit{htfl}} & \multicolumn{2}{c}{\textit{wind}} & \multicolumn{2}{c}{\textit{coast}} & \multicolumn{2}{c}{\textit{genia}} & \multicolumn{2}{c}{\textit{acl}} \\
    & Len & Cnt & Len & Cnt & Len & Cnt & Len & Cnt & Len & Cnt & Len & Cnt & Len & Cnt \\
    \midrule
    IOB sequence labeling \textit{SynTerm}             &  6 &  11 &  4 &   8 &  7 &   9 &  5 &   8 &  6 &  11 &  7 &  10 &  6 &  45 \\
    IOB sequence labeling (cross-domain)               & 14 & 806 &  9 & 1530 & 12 & 556 & 13 & 1029 & 10 & 258 & 15 & 2460 & 11 & 181 \\
    \midrule
    \code{Olmo} few-shot (cross domain)           & 11 &   0 &  8 &   3 & 13 &   3 & 13 &   2 & 11 &   3 & 13 &   4 & 16 &   3 \\
    \code{LLaMA} few-shot (cross domain)            & 27 &   0 &  8 &   3 & 14 &   2 & 13 &   3 & 11 &   1 & 15 &   4 & 18 &   6 \\
    \code{LLaMA} few-shot \textit{SynTerm}  & 15 &   2 &  10 &   2 & 17 &   1 & 14 &   2 & 13 &   1 & 15 &   4 & 18 &   7 \\
    \midrule
    \code{LLaMA} fine-tuned \textit{SynTerm}           & 15 &   3 &  9 &   2 & 15 &   3 & 14 &   2 & 14 &   3 & 14 &   4 & 18 &  10 \\
    \code{Olmo} fine-tuned \textit{SynTerm}              & 11 &   4 &  8 &   2 & 13 &   4 & 10 &   3 & 13 &   4 & 10 &   5 & 17 &  10 \\
    \midrule
    Actual               & 11 &   2 &  7 &   3 &  9 &   4 & 11 &   1 & 13 &   4 & 11 &   3 & 16 &  14 \\
    \bottomrule
  \end{tabular}
  \caption{
    Combined median term lengths (Len) and median term counts per document (Cnt) for each model and dataset.  
  }
  \label{tab:combined-median-term-stats}
\end{table*}

A possible explanation for the low performance of the few-shot models in some domains like \textit{corp} is given in  Table~\ref{tab:combined-median-term-stats}. The median number of terms extracted per document for \textit{corp} is zero for both few-shot models (with means of 0.61 and 0.82, respectively). Another contributing factor to the poor performance of the few-shot \code{LLaMA} model on \textit{corp} is the extraction of overly long terms. Table~\ref{tab:combined-median-term-stats} reveals that the median length of extracted terms is 27 characters, compared to 11 characters for gold-standard terms.

Although the IOB model trained on \textit{SynTerm} shows a large number of extracted terms per document, many extracted terms are short and imprecise, often including stopwords like “a” or “and” (see Table~\ref{tab:combined-median-term-stats}). This results in relatively high F1 scores, but with an imbalance between precision and recall (see Appendix~\ref{appendix:precision_recall}).  This imbalance and performance gap underscores the effectiveness of DiSTER and can be attributed to several key differences in model design and training paradigms. \code{XLM-R-based} models, fine-tuned for sequence labeling, are optimized for local contextual understanding within narrow task boundaries. In contrast, LLMs are trained on a broader range of tasks and contexts, equipping them with more adaptable reasoning. We posit that this flexibility allows LLMs to better leverage the \textit{SynTerm} dataset, treating it as a text generation task rather than a sequence labeling problem. Qualitative examples for these effects can be found in Appendix~\ref{appendix:qualitative_example} and an analysis of the cross‑domain instability of the sequence‑labeling models in Appendix~\ref{appenidx:instability}.

\section{Directional overlap with $k-$Nearest Neighbors}
\label{app:directional-overlap}
In order to observe the total average overlap among two datasets, we extend the analysis introduced by \citet{kambhatla2023quantifying} and include, for that, all data on both sets.
Our indicator can be defined as follows.
Let
\begin{itemize}
    \item $\mathcal{D}=\{D_1,\dots,D_N\}$ be a collection of text datasets, each embedded in $\mathbb{R}^d$;
    \item $k\in\mathbb{N}$ the fixed neighbourhood size;
    \item $N_k(x)\subset\bigcup_{i=1}^N D_i$ the (unordered) set of the $k$ nearest neighbours of sample $x$ measured w\.r.t. cosine distance in the embedding space;
    \item $\mathbf 1_{D}(y)$ the indicator that neighbour $y$ belongs to dataset $D$.
\end{itemize}
Then the Directional Overlap score $O_{A\!\to\!B}$, for two datasets $A,B\in\mathcal{D}$, is
$$
O_{A\!\to\!B}
  \;:=\;
  \frac{1}{|A|\,k}\;
  \sum_{x\in A}\;
  \sum_{y\in N_k(x)}\!
  \mathbf 1_{B}(y),
$$
\textit{i.e.} the expected fraction of a point’s $k$ closest neighbours that originate from $B$.
For convenience, one may symmetrize $O_{A\!\to\!B}$, such that
$$
O(A,B):=\tfrac12\bigl[\,O_{A\!\to\!B}+O_{B\!\to\!A}\bigr],
$$
and by construction $O(A,B)=O(B,A)\in[0,1]$ and $O(A,A)=1$.
However, in this work, we prefer the Directional Overlap indicator because it reflects the asymmetry of test and train datasets.

\begin{figure}
  \centering
\includegraphics[width=0.54\textwidth]{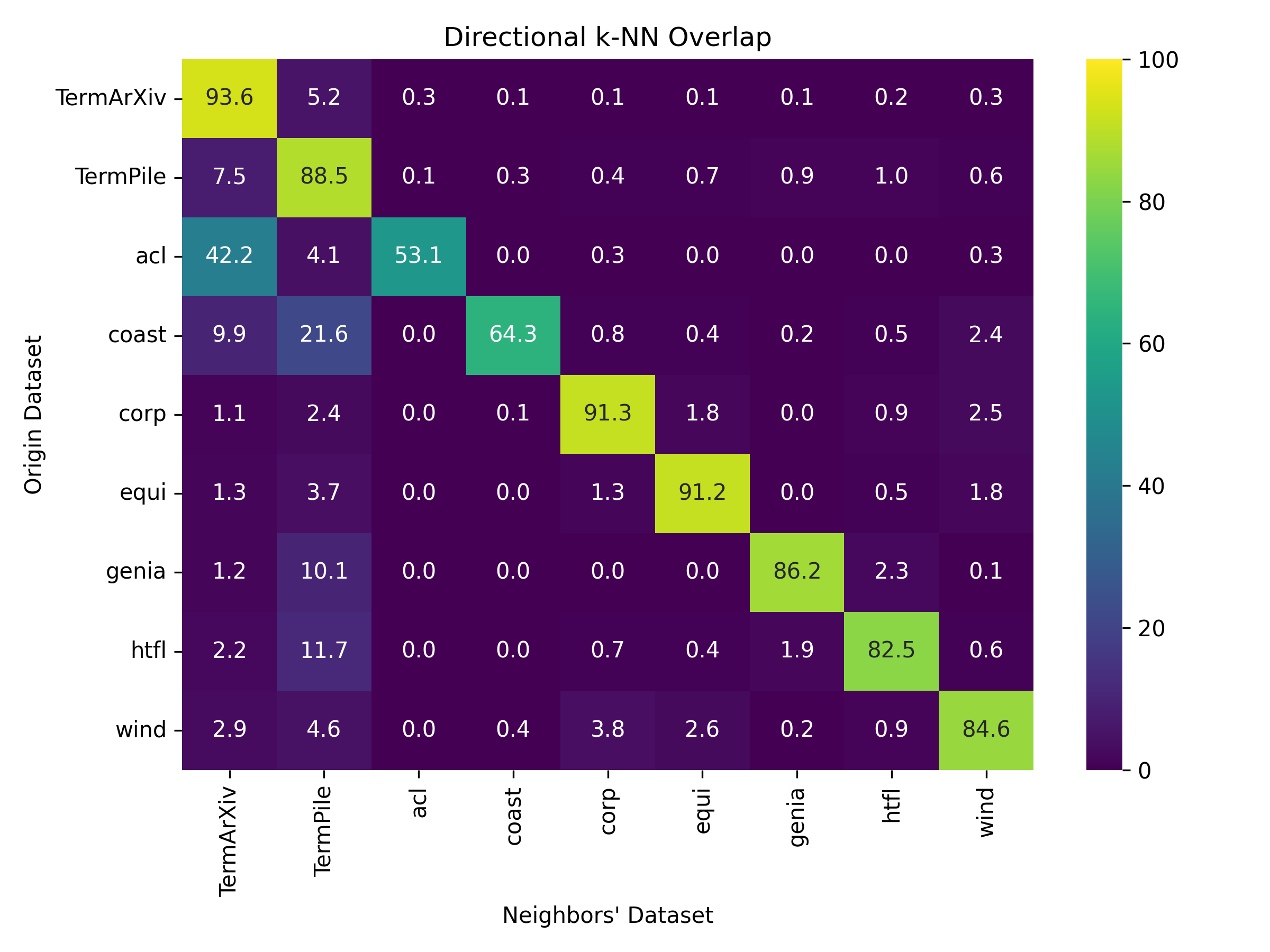}
  \caption{Directional k-NN domain overlap score.}
  \label{fig:directional-overlap}
\end{figure}

\begin{figure}
  \centering
\includegraphics[width=0.54\textwidth]{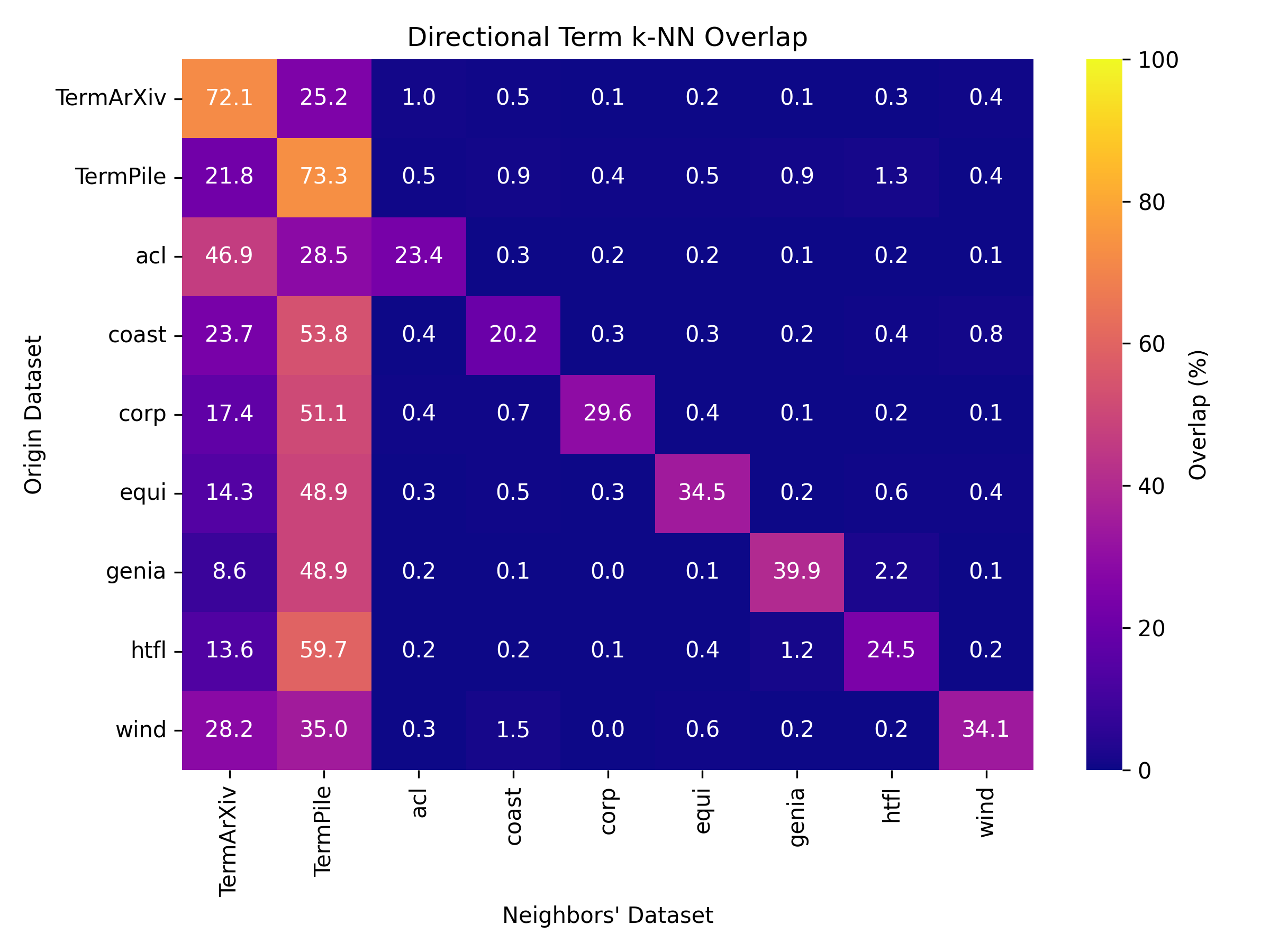}
  \caption{Directional k-NN overlap score for corpus-level terms across domains.}
  \label{fig:directional-overlap-terms}
\end{figure}
\section{Domain and Term Overlap Analysis}
\label{appendix:overlap-analysis}

To quantify dataset relationships, we extend the embedding-based Nearest Neighbors analysis from \cite{kambhatla2023quantifying} to construct a directional overlap indicator (defined in Appendix \ref{app:directional-overlap}). Figure~\ref{fig:directional-overlap} shows the percentage of each origin dataset's nearest neighbors found in each target dataset.
For interpretation, read each row as the percentage of an origin dataset's nearest neighbors found in each target dataset.

We observe that \textit{SynTerm} has measurable overlap with all test domains, illustrating the dataset’s diversity and broad coverage across different areas. 
For example for \textit{wind}, we see that 4.6 percent of its nearest neighbors come from the \textit{TermPile} dataset. By adding \textit{TermArXiv} to our \textit{SynTerm} dataset, we add more nearest neighbors for all datasets but especially data points neighboring points from \textit{acl} and \textit{coast}. 
Analyzing term-level overlap (Figure \ref{fig:directional-overlap-terms}) reveals similar patterns, with most nearest neighbors for \textit{coast} and \textit{acl} terms coming from either \textit{TermArXiv} or \textit{TermPile}. This also holds true for all other test domains to a greater or lesser extent.

These findings explain the strong performance of models fine-tuned on \textit{SynTerm} when evaluated on \textit{coast} and \textit{acl}. The effect is consistent across architectures, with the IOB-based approach gaining 46 and 42 F1 points on \textit{acl} and \textit{coast} respectively when trained on \textit{SynTerm} (Table~\ref{tab:f1_corpus}). For \textit{corp} and \textit{equi}, corpus-level term neighbors are prevalent in our synthetic datasets, but domain overlap scores are low.
This may explain why the IOB-based approach leverages \textit{SynTerm} to a lesser degree.
In order to provide a more precise picture on term overlap, we also provide in Appendix \ref{appendix:term-set-overlap} counts of unique common terms among all datasets.

Notably, \code{LLaMA} fine-tuned on \textit{TermArXiv} (containing only scientific terminology with limited overlap to most test domains) demonstrates strong transfer capabilities to unrelated domains like \textit{equi} and \textit{corp}. This suggests LLMs can learn domain-independent representations of terminology, contrasting with sequence tagging models that degrade significantly on out-of-domain data. Such findings highlight LLMs' potential to capture general principles of termhood beyond surface-level domain characteristics.

\section{Instability using supervised cross-domain Data}
\label{appenidx:instability}

While prior work on ACTER reports F1 scores above 50 on \textit{htfl} using the same IOB baseline, these results rely solely on a specific four-domain setting. Expanding to seven domains with varied validation sets reveals a sharp performance drop on \textit{htfl} (Table \ref{tab:f1_corpus} and Table \ref{tab:f1_document}). Surprisingly, increasing domain diversity using leave-one-out training does not guarantee better generalization and sometimes harms performance on individual domains. 
The directional $k$-NN overlap heat-map in Appendix \ref{appendix:overlap-analysis} confirms that \textit{htfl} exhibits only maximum 1.9\% term-level overlap with the average training mix in this case, explaining its heightened brittleness.
This highlights a key trade-off: broader training data may improve average cross-domain results but can reduce domain-specific effectiveness, especially in setups with limited resources.

Replicating the four-domain setup from the ACTER Shared Task confirms that strong results are possible, but only under specific domain splits. As shown in Table \ref{tab:IOB}, the F1 scores of the sequence labeling model vary substantially depending on which domains are used for training and validation. This instability suggests that ATE performance in prior work is highly sensitive to the choice of domains and data splits, raising concerns about the robustness and generalizability of such models. In contrast, using pseudo-labeled data from a LLM yields a more diverse and abundant training corpus, improving cross-domain robustness. Although some domain variance remains, both encoder-based and LLM-based models generalize better, underscoring the importance of data composition and not only model architecture in cross-domain ATE.

\end{document}